\documentclass{naturedraft}
\usepackage{color, graphicx, amssymb}  

\title{Universal quantum control of two-electron spin quantum bits using dynamic nuclear polarization.}

\author{Sandra Foletti$^{1 \dagger}$, Hendrik Bluhm$^{1 \dagger}$, Diana Mahalu$^2$, Vladimir Umansky$^2$ \& Amir Yacoby$^{1 \ast}$}

\begin{document}

\maketitle

\begin{affiliations}
 \item Department of Physics, Harvard University, Cambridge, MA 02138, USA
 \item Braun Center for Submicron Research, Department of Condensed Matter Physics, Weizmann Institute of Science, Rehovot 76100, Israel
 \item [$\dagger$] These authors contributed equally to this work.
 \item [$\ast$] e-mail: yacoby@physics.harvard.edu
\end{affiliations}

\begin{abstract} 
One fundamental requirement for quantum computation is to perform universal manipulations of quantum bits at rates much faster than the qubit's rate of decoherence. 
Recently, fast gate operations have been demonstrated in logical spin qubits composed of two electron spins where the rapid exchange of the two electrons permits electrically controllable rotations around one axis of the qubit. However, universal control of the qubit requires arbitrary rotations around at least two axes. 
Here we show that by subjecting each electron spin to a magnetic field of different magnitude we achieve full quantum control of the two-electron logical spin qubit with nanosecond operation times.  Using a single device, a magnetic field gradient of several hundred milliTesla is generated and sustained using dynamic nuclear polarization of the underlying Ga and As nuclei. Universal control of the two-electron qubit is then demonstrated using quantum state tomography. 
The presented technique provides the basis for single and potentially multiple qubit operations with gate times that approach the threshold required for quantum error correction.

\end{abstract}

\begin{figure}
\begin{center}
\includegraphics[width=8.9cm]{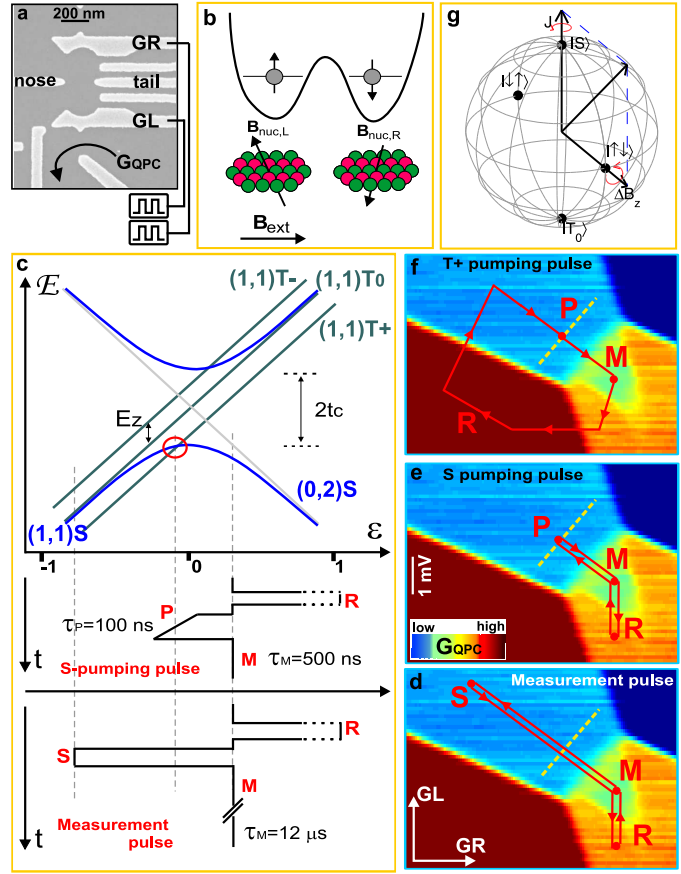}
\end{center}
\caption{\textbf{Pump and measurement schemes.} \textbf{a,} Geometrical representation (Bloch sphere) of the two level system ($S$ and $T_0$) and the two rotation axes ($J$ and $\Delta B_{nuc}^z$) allowing the implementation of universal single qubit gates. \textbf{b,} SEM micrograph of a device similar to the one measured. Gates GL and GR control the charge configuration of the two dots, the central gates (Nose and Tail) control the tunneling rate between the two dots. The average charge configuration is detected by measuring the conductance ($G_{QPC}$) through a capacitively coupled quantum point contact. \textbf{c,} $\mathbf{B_{nuc,L}}$ and $\mathbf{B_{nuc,R}}$ are the local magnetic fields experienced by the electrons in the left and right dot through hyperfine coupling with the Ga and As nuclei. \textbf{d,} Schematic representation of the energy levels at the (0,2)-(1,1) charge transition for finite external magnetic field. The detuning $\varepsilon$ from the degeneracy point is controlled by the voltages on GL and GR. Two pulse cycle are presented: 1) Nuclear pumping: the system is moved to point P where $S$ and $T_+$ are degenerate and can mix 2) Measurement pulse: the system is moved to large negative detuning where the states $S$ and $T_0$ can mix. \textbf{e,} The measurement pulse scheme \textbf{f,} The $S$-pumping pulse scheme \textbf{g,} The $T_+$-pumping pulse scheme, all shown as a function of GL and GR. }
\label{Fig1}
\end{figure}

The potential realization of quantum computers has attracted a lot of attention because of their promise to perform certain calculations practically intractable for classical computers. While a classical bit attains only two values (0 and 1), 
the phase space of a quantum bit (a two-level system) is in one-to-one correspondence with the points on the surface of a three dimensional sphere, known as the Bloch sphere\cite{Chuang2000}, where the basis states (corresponding to the classical 0 and 1) are represented at the north and south pole (Fig. \ref{Fig1}a).
A generic manipulation of the qubit needed to implement universal gate operations requires the ability to perform rotations around two axes in the Bloch sphere\cite{Levy2002, Coish2007, Press2008, Hanson2007} (for example the $z$ and $x$-axis).
In the present work, the two-level quantum bit (smallest logical unit of the quantum computer) is encoded in the spin state of two electrons confined in a double-well potential.
This semiconductor-based system has potential for good scalability, manipulations are all-electrical and potentially fast enough to enable $10^4$ universal gate operations within the coherence time, an essential requirement for quantum error correction\cite{Aharonov1999}. 
 
For the two-electron spin qubits, rotations around the $z$-axis, corresponding to the coherent exchange of two electrons, have recently been demonstrated by Petta \textit{et al.}\cite{Petta2005}. 
Rotations around the second axis require the presence of a non-uniform magnetic field across the double-well potential, making the two spins precess at different rates. Here we take advantage of the interaction of the electrons with the nuclear  magnetic field of the Ga and As sublattices of the host material in order to generate the required magnetic field gradient. While fluctuations of this hyperfine field are known to be a major source of decoherence\cite{Khaetskii2002,Merkulov2002, Witzel2006, Koppens2005, Zhang2006},
in this letter we demonstrate the possibility of building up a gradient in the hyperfine field that significantly exceeds the fluctuations and can be sustained for times longer than 30 min.
This is done by employing pumping schemes that transfer spin and thus magnetic moment from the electronic system to the nuclei.
Internally created gradients of nuclear field, in excess of 200~mT, together with the coherent exchange of the two electrons allow us to rapidly manipulate the two-electron spin qubit.  
The coherent manipulation is demonstrated by reconstructing the evolution of the state within the Bloch sphere through quantum state tomography.

The double-well potential that confines the electrons is formed by applying a negative voltage to metal gates deposited on top of a two dimensional electron gas embedded in a GaAs/AlGaAs heterostructure. The negative potential depletes the electrons underneath the metal gates creating two isolated puddles of electrons (double quantum dot, Fig. \ref{Fig1}b). The number of electrons in the dots can be controlled by tuning the potentials on the gates. 
We restrict the total occupation of the double quantum dot to two electrons, and describe their spatial separation by the parameter $\varepsilon$: for $\varepsilon \gg 0$ both electrons are in the right quantum dot, the (0,2) configuration; for $\varepsilon \ll 0$ one electron occupies each dot, the (1,1) configuration. The parameter $\varepsilon$ and hence the dots' charge state can be continuously swept through intermediate configurations by varying the voltages on the metal gates.
In the (0,2) charge configuration, the only energetically accessible spin configuration is the singlet state  $S$(0,2) $=(\uparrow\downarrow \!\!-\!\!\downarrow\uparrow)/\sqrt{2}$ (the arrows indicate the direction of the electron spins).
As we separate the electrons, the wavefunctions overlap decreases and four spin configurations become energetically degenerate:  the singlet $S$(1,1) and three triplets  $T_0 =(\uparrow\downarrow \!\!+\!\! \downarrow\uparrow)/\sqrt{2}$,  $T_-=\downarrow\downarrow$ and  $T_+=\uparrow\uparrow$ (Fig. \ref{Fig1}d).
We select the states $S$(1,1) and $T_0$, both having zero $z$ component of the spin angular momentum, as the basis states of our logical qubit\cite{Lidar1998, Taylor2005} and lift the degeneracy with the states $T_-$ and $T_+$ by applying an external magnetic field $B_{ext}$. The Zeeman energy $E_z=g\mu_B B$ ($g=-0.4$ is the $g$-factor for GaAs, $\mu_B$ Bohr's magneton) shifts the $T_+$ state to lower energies, creating a crossing point with the singlet (marked by a red circle in Fig. \ref{Fig1}d) at a value of $\varepsilon$ that depends on $\mathbf{B}=\mathbf{B_{ext}}+ \mathbf{B_{nuc}}$, where $\mathbf{B_{nuc}} \equiv (\mathbf{B_{nuc,L}}+\mathbf{B_{nuc,R}})/2$ is the average hyperfine field and $\mathbf{B_{nuc,L}}$ and $\mathbf{B_{nuc,R}}$ the nuclear fields felt by the electron in the left and right dot, respectively (Fig. \ref{Fig1}c). 
Within this logical subspace, rotations around the $z$-axis are controlled by the energy splitting between $S$(1,1) and $T_0$, denoted by $J(\varepsilon)$. 
This evolution amounts to a coherent exchange of the two electrons. 
Rotations around the $x$-axis are controlled by the $z$-component of a magnetic field gradient across the two electrons, $\mathbf{\Delta B_{nuc}}=\mathbf{B_{nuc, L}}-\mathbf{B_{nuc,R}}$.  
If we thus let a state evolve around a combined axis $J\textbf{z}+g \mu_B \Delta B_{nuc}^z\textbf{x}$, the precession frequency will be given by $f=\sqrt{J^2+(g\mu_B \Delta B_{nuc}^z)^2}/h$ ($h \approx 4 \cdot 10^{-15}$ $ eV \cdot s$ is Plank's constant).

While controlled $z$-rotations have been previously shown\cite{Petta2005}, controlled rotations around the $x$-axis of the two-electron logical qubit have not been demonstrated to date. Clearly, the challenge is to provide a stable magnetic field gradient across the two dots which exceeds the intrinsic nuclear fluctuations due to the hyperfine interaction.
Here we present two polarization schemes by which the gradient can be increased to values significantly exceeding its fluctuations. Both pumping schemes make use of the degeneracy point between $S$(1,1) and $T_+$. Transitions between the two states that are driven by the transverse component of $\mathbf{\Delta B_{nuc}}$\cite{Taylor2007} are accompanied by a spin flip of the nuclei in order to conserve the total angular momentum.
Our first pumping scheme follows a standard recipe\cite{Petta2008, Reilly2008C} of initializing the system in the $S$(0,2) state followed by a 50 or 100 ns long sweep across the $S$--$T_+$ degeneracy point. This process ideally transfers one unit of angular momentum into the nuclear sub system. 
In addition, we have developed an alternative pumping scheme whereby we initialize the system in a $T_+$(1,1) state followed by a similar slow passage through the $S$--$T_+$ degeneracy point. This new $T_+$-pumping scheme allows us to polarize the nuclear subsystem in a direction opposite to the $S$-pumping scheme. The $T_+$-pumping scheme works only when the Zeeman energy exceeds the electron temperature in the reservoirs: 
the system is swept slowly into (0,1) and subsequently reloaded into the (1,1) charge state (Fig. \ref{Fig1}g).
First the right and then the left electron align with the external field due to large Zeeman energy ($\approx$ 12.5 $\mu eV$ at 500 mT), which preferentially loads a $T_+$ state.

While the above nuclear pumping schemes should produce nuclear polarization, it is not obvious at all that this nuclear polarization should be different across the two dots\cite{Reilly2008C}. Since the mixing between the $S$(1,1) and $T_0$(1,1) is only sensitive to the field gradient, we use a pulse cycle that monitors the coherent evolution around the $x$-axis in order to measure this gradient.
The system is first reset into a $S$(0,2) state. $\varepsilon$ is then abruptly set to point \textbf{S} in (1,1) for an evolution time $\tau_S$ (see Fig. \ref{Fig1}e). Here $\Delta B_{nuc}^z$ $\gg J(\varepsilon)/g\mu_B$ drives coherent oscillations between $S$(1,1) and $T_0$ and the probability of being in a singlet state oscillates in time as $p(S)=\cos^2(g \mu_B \Delta B_{nuc}^z \cdot \tau_S/2 \hbar)$.
When the system is brought back to the measurement point \textbf{M} only transitions from $S$(1,1) to $S$(0,2) are allowed, while $T_0$ remains blocked in the (1,1) charge configuration. This spin-blockade effect allows to map the spin configuration of the state onto a charge configuration\cite{Johnson2005}, which is measured by a charge sensor\cite{Johnson2005}. Here we use a quantum point contact (QPC) positioned next to the double quantum dot (Fig. \ref{Fig1}b)
in order to detect changes in the double dot charge configuration. 
The QPC signal, averaged over many gradient-probing cycles, is proportional to the probability of being in a singlet state.

\begin{figure}
\begin{center}
\includegraphics[width=17.8cm]{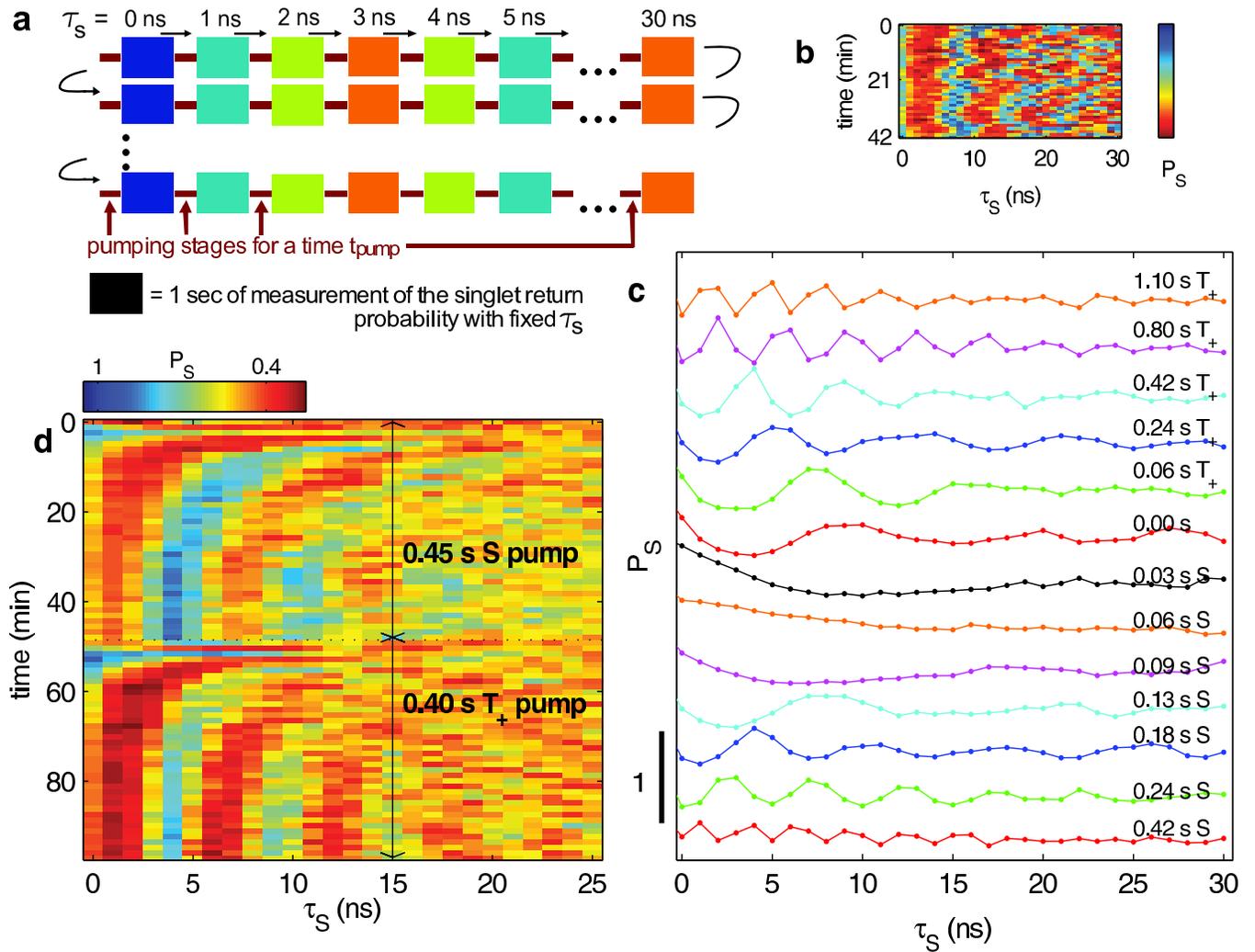}
\end{center}
\caption{\textbf{Build-up of a gradient with two different pumping cycles.} 
\textbf{a,} Schematics explaining the measurement of the singlet return probability as a function of $\tau_S$. Each measurement point (rectangle) is an average over 1 sec and preceded by a pump cycle that runs for a time $t_{pump}$. Each line is a repetition of the same measurement procedure. 
\textbf{b,} Measurement of the singlet return probability as a function of $\tau_S$ for $t_{pump}=$ 60 ms. 
\textbf{c,} Singlet return probability as a function of separation time $\tau_S$ for different $t_{pump}$. Each data set line is an average over 30 $\tau_S$ sweeps. $P_s$ is normalized by the size of the DC charge transition (see Supplementary Material). Traces are displaced for clarity.
\textbf{d,} Crossover between $S$ and $T_+$-pumping. The plot shows an average of nine subsequent repetitions of the same measurement. The disappearance of the oscillations upon changing the pump pulse followed by a recovery (around 0 and 50 min) suggests that  $S$ and $T_+$-pumping produce $\Delta B_{nuc}^z$ of opposite sign. $B_{ext}=1.5$ T for these data sets.}
\label{Fig2}
\end{figure}

A steady state nuclear field can be achieved by continuously alternating between a pump cycle that runs for a time $t_{pump}$, and a gradient probing cycle that runs for 1 sec, as schematically visualized in Fig.~\ref{Fig2}a.
In each measurement stage a gradient probing pulse with a different separation time $\tau_S$ is used 
(0 $\le\!\!\tau_S \!\!\le$ 30 ns for each $\tau_S$-sweep). The outcome of a measurement repeating 
$\tau_S$-sweeps 40 times and using $t_{pump}=$ 60 ms is shown in Fig. \ref{Fig2}b. An oscillatory signal with a frequency fluctuating around a steady mean is clearly visible. 
In the present measurement the gradient is kept in a steady state for 40 minutes, but this time could have been extended indefinitely. 
Each curve in Fig. \ref{Fig2}c shows an average over 30 $\tau_S$ sweeps and the different values of
$t_{pump}$ control the steady state value of the gradient in each data set.   
We observe that the oscillations vanish (corresponding to a $\Delta B_{nuc}^z$ fluctuating around 0) at moderate $S$-pumping rather then $t_{pump}$ = 0. This appears to reflect a small polarization effect from the measurement pulses that can be compensated with $S$-pumping (see Supplementary Material).
To compare the $S$ and $T_+$-pumping schemes, we have taken a measurement where we have switched between $S$-pumping and $T_+$-pumping every 40 $\tau_s$ sweeps (see Fig. \ref{Fig2}c). The data show that upon changing the pump cycle, the oscillations disappear and then recover after a few minutes in a way that suggests a sign change of the induced gradient.

\begin{figure}
\begin{center}
\includegraphics[width=8.9cm]{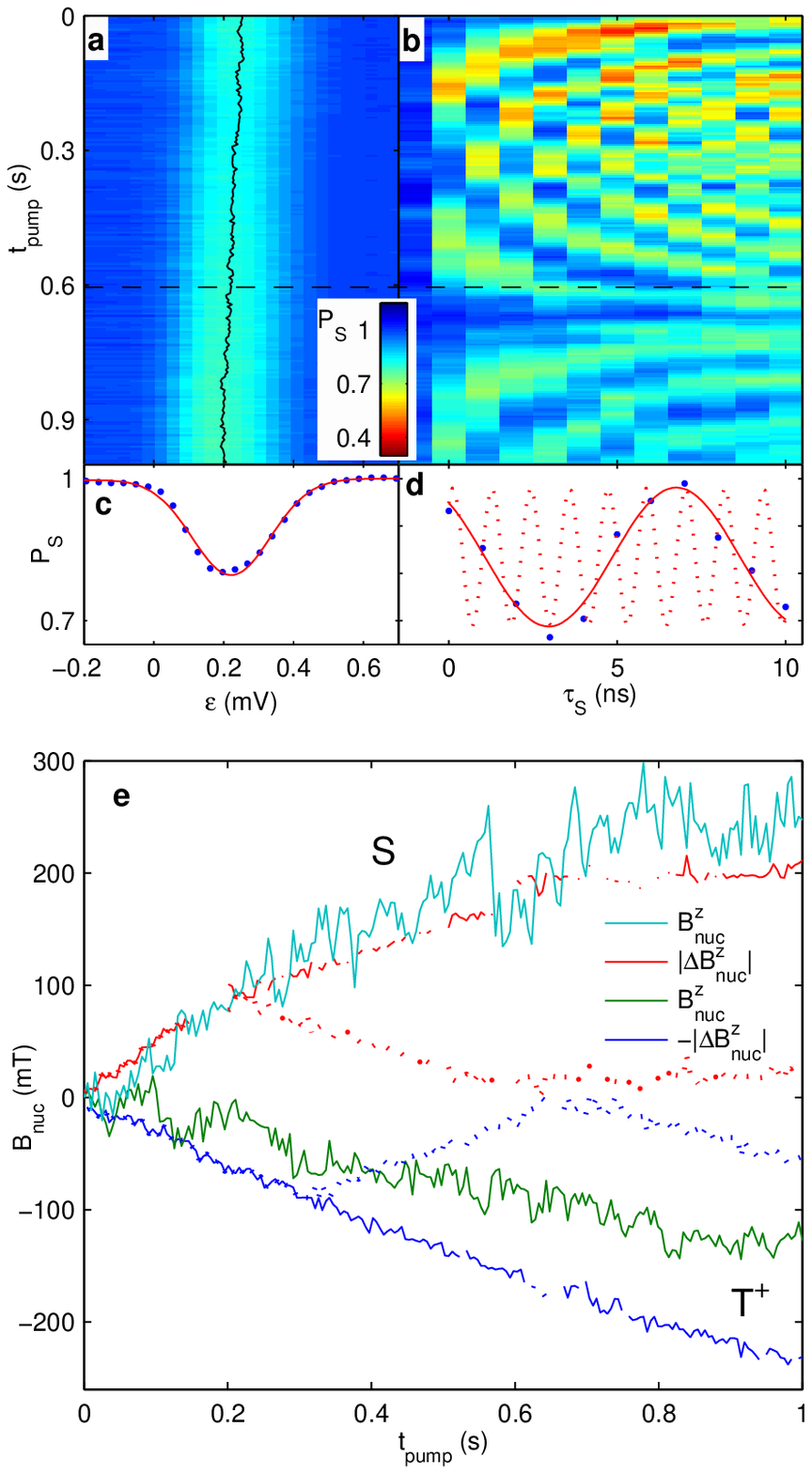}
\end{center}
\caption{\textbf{Comparison of the gradient and the average nuclear field}. \textbf{a,} Line scans of the position of the $S$--$T_+$ transition as a function of pump time: the $T_+$-pumping cycle, applied for a duration $t_{pump}$, precedes each 2 s long measurement interval. \textbf{b,} Singlet return probability versus $\tau_S$ under the same pumping conditions as in \textbf{a}. The repeated increase and decrease of the apparent oscillation frequency is a consequence of aliasing of a monotonically increasing frequency due to the 1 ns sampling interval. \textbf{c,} Single line scan from the data in panel \textbf{a} at $t_{pump}=0.6 $ s. \textbf{d,} The corresponding $\tau_S$ sweep from panel \textbf{b}. The continuous red line is a sinusoidal fit, the dashed line reconstructs the actual, non-aliased time dependence of the singlet probability.  \textbf{e,} $B_{nuc}$ and $\Delta B_{nuc}^z$ extracted from fits as shown in \textbf{c}, \textbf{d} as a function of $t_{pump}$ for both $T_+$ and $S$-pumping. The shift of the $S$--$T_+$ transition was converted to $B_{nuc}^z$ using its measured dependence on $B_{ext}$ (see Supplementary Material) \cite{Petta2008}, $\Delta B_{nuc}^z$ was obtained from the fitted oscillation frequency corrected for aliasing. The dotted lines shows the same data before this correction. The fluctuations in the $B_{nuc}^z$ curves reflect measurement noise. $B_{ext}=500$ mT for these data sets.}
\label{Fig3}
\end{figure}

While the magnitude of the gradient is determined via coherent $x$-rotations, we can also measure the average value of the nuclear field by monitoring the position of the $S$--$T_+$ transition\cite{Petta2008}.
This should clarify whether spins are flipped only in one or both dots. 
Figs. \ref{Fig3}a,b show interleaved measurements of the position of the $S$--$T_+$ transition\cite{Reilly2008C} and the oscillatory $S$--$T_0$ mixing as a function of $t_{pump}$ using the $T_+$-pumping cycle.
A shift of the $S$--$T_+$ transition to more negative $\varepsilon$ corresponds to the build-up of an average field $B_{nuc}^z$ oriented opposite to the external magnetic field, consistent with spin flips from down to up in the nuclear system\cite{Paget1977}.
Fig. \ref{Fig3}e shows that at $B_{ext}$= 500 mT, $\Delta B_{nuc}^z$ reaches 230 mT while $B_{nuc}^z$ is about 130 mT. The ratio of nearly a factor 2 indicates that the nuclei are polarized predominantly in one of the two dots.
Data obtained using the $S$-pumping cycle (Fig. \ref{Fig3}e) show a $\Delta B_{nuc}^z$ that tends to be slightly smaller than the average field. The value of $B_{nuc}^z$ can be subject to various systematic analysis errors, but it is clearly not much larger than $\Delta B_{nuc}^z$ (see Supplementary Material).

\begin{figure}
\begin{center}
\includegraphics[width=18.3cm]{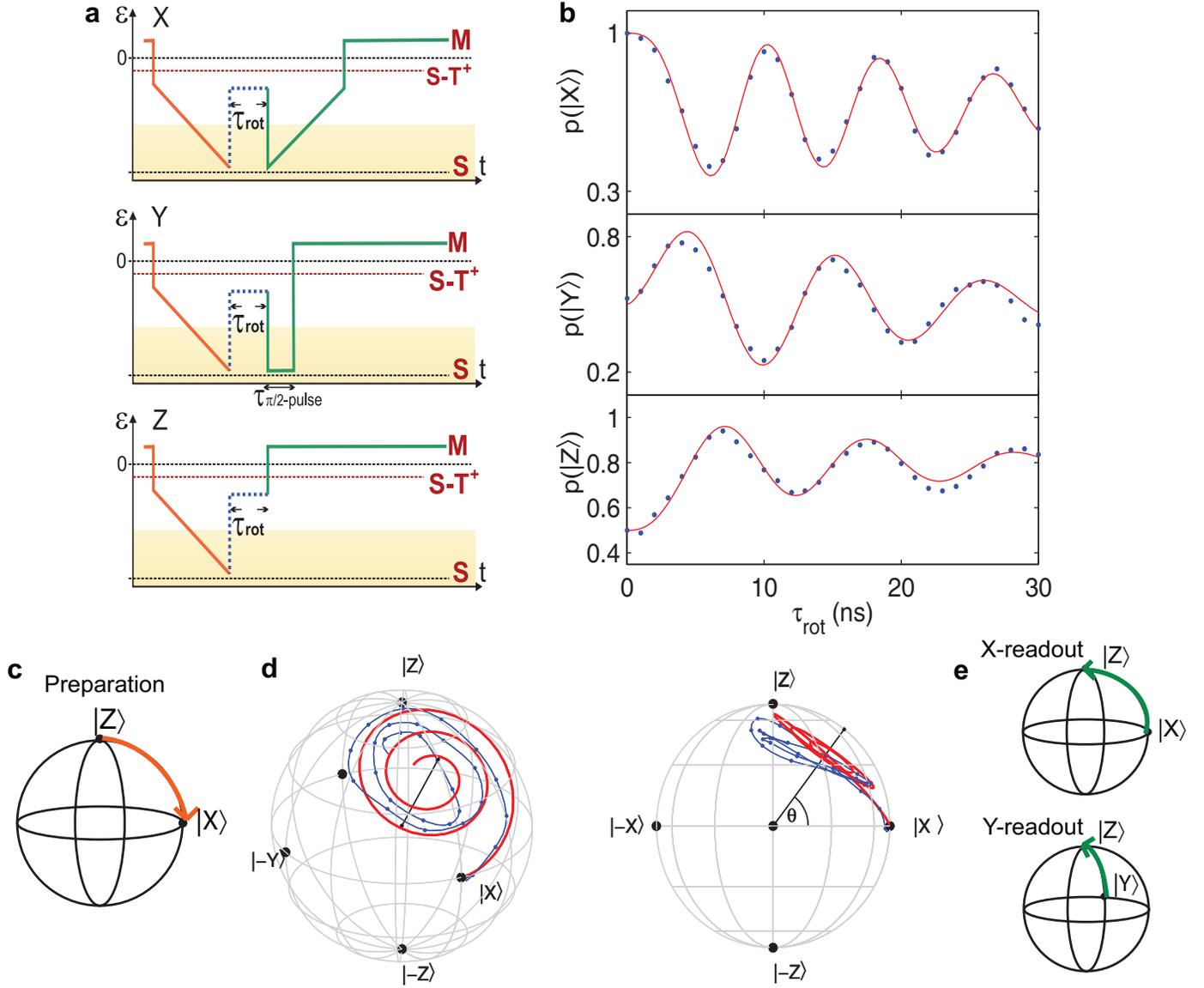}
\end{center}
\caption{\textbf{State tomography and universal gate.}
  \textbf{a,} Pulse schemes to measure the singlet probability $p(|S\rangle)\equiv p(|Z\rangle)$ , the $|\!\!\uparrow \downarrow \rangle$ probability $p(|\!\!\uparrow \downarrow \rangle)\equiv p(|X\rangle)$ and the $|S\rangle +i|T_0\rangle$ probability $p(|S\rangle +i|T_0\rangle) \equiv p(|Y\rangle)$ after rotation around a tilted, tunable  axis. \textbf{b,} Measurements taken  with the X, Y and Z pulses (dots) and fits (line) to a numerical solution of the Schr\"odinger equation of the $S$--$T_0$ Hamiltonian incorporating the finite pulse rise time and inhomogeneous broadening due to fluctuations in $\Delta B_{nuc}^z$. 
\textbf{c,} An $|\!\!\uparrow\downarrow \rangle$=$|X \rangle$ state is prepared by adiabatically turning off $J$. 
\textbf{d,} Representation of the measured and fitted trajectories in the Bloch sphere. In order to eliminate phase shifts due to the slightly different frequencies (see text), the time scales for the X-data has been rescaled using spline interpolation so that the expected phase relations are maintained. The blue line is a spline interpolation of the data points in panel \textbf{b}.
\textbf{e,} Visualization of the X and Y-readout schemes on the Bloch sphere. 
}
\label{Fig4}
\end{figure}

Combining our slowly tunable $x$-rotation gate with the electrically controllable exchange gate allows single qubit rotations around an axis that can be rapidly tilted to any desired angle between 0 and nearly $\pi/2$ away from the $x$-axes (angle $\theta$ in Fig. \ref{Fig4}d).
Concatenating rotations around different axis allows to implement universal quantum control. 
We demonstrate and characterize the rotation around an arbitrary axes using state tomography, consisting of three independent measurements of the probability of being in a $|S\rangle \equiv |Z\rangle$, in an $|S\rangle$+$|T_0\rangle \equiv |\!\!\uparrow\downarrow\rangle \equiv |X\rangle$ and in a $|S\rangle$+$i|T_0\rangle \equiv |Y\rangle$ state\cite{Chuang2000}, with pulses shown in Fig. \ref{Fig4}a. 
This allows us to fully reconstruct the time evolution of the state vector.
For each of the measurements, we first prepare an $|\!\!\uparrow\downarrow\rangle$ state by loading a $S$(0,2) and adiabatically switching off $J(\varepsilon)$ in (1,1). The desired rotation is performed by quickly setting $J$ to a finite value for a time $\tau_{rot}$.
Rapidly returning to $S(0,2)$ allows to measure $p(|Z\rangle) \equiv |\langle Z|\psi\rangle|^2$, whereas slowly increasing $J$ brings $|\!\!\uparrow\downarrow\rangle$ onto $|S\rangle$ and $|\!\!\downarrow\uparrow\rangle$ onto $|T_0\rangle$, thus allowing the readout of $p(|X \rangle) \equiv |\langle X|\psi\rangle|^2$.
To obtain $p(|Y \rangle) \equiv |(\langle Y|)|\psi\rangle|^2$, $J$ is turned off for a time corresponding to a $\pi/2$ rotation around the $x$-axis before rapidly returning to \textbf{M}.
Results of this procedure for a particular choice of $J$ and $\Delta B_{nuc}^z$ are shown in Fig. \ref{Fig4}c as a function of $\tau_{rot}$.
For ideal pulses, one would expect $p(|X \rangle)$ to oscillate sinusoidally between 1 and $(1+\cos(2\theta))/2$, $p(|Z \rangle)$ between 1/2 and $(1+\sin(2\theta))/2$, whereas $p(|Y \rangle )$ should vary symmetrically around 1/2.
Deviations from this behavior can be attributed to a finite pulse rise time and high pass filtering of the pulses.
The first causes an approximately adiabatic drift of the rotation axis which prevents $p(|Z \rangle)$ to return to the starting point, whereas the second leads to slightly different $\varepsilon$-offsets and thus different $J(\varepsilon)$ for different pulses, causing a $\approx$ 25~\% frequency change.
Fits to a model (Fig. \ref{Fig4}b, see Supplementary Material) incorporating these effects and inhomogeneous broadening due to fluctuations in $\Delta B_{nuc}^z$ give a good match with the data. In Fig. \ref{Fig4}d, the data and fits are displayed in the Bloch sphere representation.
We estimate that the errors due to measurement noise,  pulse imperfections other than those included in the model, incomplete ensemble averaging over nuclear fluctuations and uncertainties in the QPC conductance calibration (Supplementary Material) are on the order of 0.15 for all three probability measurements. They could be substantially reduced by improving the characteristics of our high frequency setup, such that pulse compensation schemes would be simpler and pulse dependent variations of $J$ could be eliminated.

The mechanism responsible for the large gradient due to pumping is currently unknown. 
One possible cause is an asymmetry in the size of the two dots due to local disorder. Both the probability to flip a nuclear spin in one of the dots and the change in hyperfine field due to that flip are inversely proportional to the number of nuclei N over which the electron wave function extends
\cite{Stopa2009}. The overall $1/N^2$ dependance results in the smaller dot being polarized more rapidly. Different relaxation rates in the two dots and more complex aspects of the nuclear dynamics may also play a role\cite{Maletinsky2007}. 
The relation between our results and those in Ref. [17], where a strong suppression of $\Delta B_{nuc}^z$ was reported, is currently under investigation. The apparent contradiction with the observation of Reilly \emph{et al.}\cite{Reilly2008B} that $S$-pumping becomes ineffective at fields exceeding a few tens of mT might be due to a different coupling between the dots and to the electron reservoirs (see Supplementary Material).

We demonstrated  the ability to perform universal single qubit operations in sub nanosecond time scales\cite{Greilich2009}, two orders of magnitude faster than previously shown for single spin qubits\cite{Koppens2006, Nowack2007, Pioro2008}. These short operation times together with the demonstrated coherence times of a few microseconds\cite{Petta2005} and predicted coherence times of up to 100 microseconds\cite{Yao2006, Cywinski2009B, Coish2008} suggest that the requirements for quantum error correction of two-electron spin qubits are within reach. Furthermore, our ability to record the magnetic field gradient opens the way towards feedback control of the nuclear environment that would prolong $T_2^*$ and thereby reduce the number of error correcting pulses needed.

\bibliography{../../bibdata,bibdata}

\begin{thebibliography}{10}
\expandafter\ifx\csname url\endcsname\relax
  \def\url#1{\texttt{#1}}\fi
\expandafter\ifx\csname urlprefix\endcsname\relax\def\urlprefix{URL }\fi
\providecommand{\bibinfo}[2]{#2}
\providecommand{\eprint}[2][]{\url{#2}}

\bibitem{Chuang2000}
\bibinfo{author}{Nielsen, M.~A.} \& \bibinfo{author}{Chuang, I.~L.}
\newblock \emph{\bibinfo{title}{Quantum Computation and Quantum Information}}
  (\bibinfo{publisher}{Cambridge University Press}, \bibinfo{year}{2000}).

\bibitem{Levy2002}
\bibinfo{author}{Levy, J.}
\newblock \bibinfo{title}{Universal quantum computation with spin-1/2 pairs and
  heisenberg exchange}.
\newblock \emph{\bibinfo{journal}{Phys.\ Rev.\ Lett.}}
  \textbf{\bibinfo{volume}{89}}, \bibinfo{pages}{147902}
  (\bibinfo{year}{2002}).

\bibitem{Coish2007}
\bibinfo{author}{Coish, W.~A.} \& \bibinfo{author}{Loss, D.}
\newblock \bibinfo{title}{Exchange-controlled single-electron-spin rotations in
  quantum dots}.
\newblock \emph{\bibinfo{journal}{Phys.\ Rev.\ B}}
  \textbf{\bibinfo{volume}{75}}, \bibinfo{pages}{161302}
  (\bibinfo{year}{2007}).

\bibitem{Press2008}
\bibinfo{author}{Press, D.}, \bibinfo{author}{Ladd, T.},
  \bibinfo{author}{Zhang, B.} \& \bibinfo{author}{Yamamoto, Y.}
\newblock \bibinfo{title}{Complete quantum control of a single quantum dot spin
  using ultrafast optical pulses}.
\newblock \emph{\bibinfo{journal}{Nature}} \textbf{\bibinfo{volume}{456}},
  \bibinfo{pages}{218} (\bibinfo{year}{2008}).

\bibitem{Hanson2007}
\bibinfo{author}{Hanson, R.} \& \bibinfo{author}{Burkard, G.}
\newblock \bibinfo{title}{Universal set of quantum gates for double-dot spin
  qubits with fixed interdot coupling}.
\newblock \emph{\bibinfo{journal}{Phys.\ Rev.\ Lett.}}
  \textbf{\bibinfo{volume}{98}}, \bibinfo{pages}{050502}
  (\bibinfo{year}{2007}).

\bibitem{Aharonov1999}
\bibinfo{author}{Aharonov, D.} \& \bibinfo{author}{Ben-Or, M.}
\newblock \bibinfo{title}{Fault-tolerant quantum computation with constant
  error rate} (\bibinfo{year}{1999}).
\newblock \eprint{arXiv:cond-mat/9906129}.

\bibitem{Petta2005}
\bibinfo{author}{Petta, J.~R.} \emph{et~al.}
\newblock \bibinfo{title}{Coherent manipulation of coupled electron spins in
  semiconductor quantum dots}.
\newblock \emph{\bibinfo{journal}{Science}} \textbf{\bibinfo{volume}{309}},
  \bibinfo{pages}{2180} (\bibinfo{year}{2005}).

\bibitem{Khaetskii2002}
\bibinfo{author}{Khaetskii, A.~V.}, \bibinfo{author}{Loss, D.} \&
  \bibinfo{author}{Glazman, L.}
\newblock \bibinfo{title}{Electron spin decoherence in quantum dots due to
  interaction with nuclei}.
\newblock \emph{\bibinfo{journal}{Phys.\ Rev.\ Lett.}}
  \textbf{\bibinfo{volume}{88}}, \bibinfo{pages}{186802}
  (\bibinfo{year}{2002}).

\bibitem{Merkulov2002}
\bibinfo{author}{Merkulov, I.~A.}, \bibinfo{author}{Efros, A.~L.} \&
  \bibinfo{author}{M.Rosen}.
\newblock \bibinfo{title}{Electron spin relaxation by nuclei in semiconductor
  quantum dots}.
\newblock \emph{\bibinfo{journal}{Phys. Rev. B}} \textbf{\bibinfo{volume}{65}},
  \bibinfo{pages}{205309} (\bibinfo{year}{2002}).

\bibitem{Witzel2006}
\bibinfo{author}{Witzel, W.~M.} \& \bibinfo{author}{{Das Sarma}, S.}
\newblock \bibinfo{title}{Quantum theory for electron spin decoherence induced
  by nuclear spin dynamics in semiconductor quantum computer architecture}.
\newblock \emph{\bibinfo{journal}{Phys.\ Rev.\ B.}}
  \textbf{\bibinfo{volume}{74}}, \bibinfo{pages}{035322}
  (\bibinfo{year}{2006}).

\bibitem{Koppens2005}
\bibinfo{author}{{Koppens}, F. H.~L.} \emph{et~al.}
\newblock \bibinfo{title}{Control and detection of singlet-triplet mixing in a
  random nuclear field}.
\newblock \emph{\bibinfo{journal}{Science}} \textbf{\bibinfo{volume}{309}},
  \bibinfo{pages}{1346} (\bibinfo{year}{2005}).

\bibitem{Zhang2006}
\bibinfo{author}{Zhang, W.}, \bibinfo{author}{Dobrovitki, V.},
  \bibinfo{author}{Al-Hassanieh, K.}, \bibinfo{author}{Dagotto, E.} \&
  \bibinfo{author}{Harmon, B.}
\newblock \bibinfo{title}{Hyperfine interaction induced decoherence of electron
  spins in quantum dots}.
\newblock \emph{\bibinfo{journal}{Phys. Rev. B}} \textbf{\bibinfo{volume}{74}},
  \bibinfo{pages}{205313} (\bibinfo{year}{2006}).

\bibitem{Lidar1998}
\bibinfo{author}{Lidar, D.~A.}, \bibinfo{author}{Chuang, I.~L.} \&
  \bibinfo{author}{Whaley, K.~B.}
\newblock \bibinfo{title}{Decoherence-free subspace for quantum computation}.
\newblock \emph{\bibinfo{journal}{Phys.\ Rev.\ Lett.}}
  \textbf{\bibinfo{volume}{81}}, \bibinfo{pages}{2594} (\bibinfo{year}{1998}).

\bibitem{Taylor2005}
\bibinfo{author}{Taylor, J.~M.} \emph{et~al.}
\newblock \bibinfo{title}{Fault-tollerant architecture for quantum computation
  using electrically controlled semiconductor spins}.
\newblock \emph{\bibinfo{journal}{Nature Physics}}
  \textbf{\bibinfo{volume}{1}}, \bibinfo{pages}{177} (\bibinfo{year}{2005}).

\bibitem{Taylor2007}
\bibinfo{author}{Taylor, J.~M.} \emph{et~al.}
\newblock \bibinfo{title}{Relaxation, dephasing, and quantum control of
  electron spins in double quantum dots}.
\newblock \emph{\bibinfo{journal}{Phys.\ Rev.\ B.}}
  \textbf{\bibinfo{volume}{76}}, \bibinfo{pages}{035315}
  (\bibinfo{year}{2007}).

\bibitem{Petta2008}
\bibinfo{author}{Petta, J.~R.} \emph{et~al.}
\newblock \bibinfo{title}{Dynamic nuclear polarization with single electron
  spins}.
\newblock \emph{\bibinfo{journal}{Phys.\ Rev.\ Lett.}}
  \textbf{\bibinfo{volume}{100}}, \bibinfo{pages}{067601}
  (\bibinfo{year}{2008}).

\bibitem{Reilly2008C}
\bibinfo{author}{Reilly, D.~J.} \emph{et~al.}
\newblock \bibinfo{title}{Suppressing spin qubit dephasing by nuclear state
  preparation}.
\newblock \emph{\bibinfo{journal}{Science}} \textbf{\bibinfo{volume}{321}},
  \bibinfo{pages}{817} (\bibinfo{year}{2008}).

\bibitem{Johnson2005}
\bibinfo{author}{{Johnson}, A.~C.} \emph{et~al.}
\newblock \bibinfo{title}{Triple-singlet spin relaxation via nuclei in a double
  quantum dot}.
\newblock \emph{\bibinfo{journal}{Nature}} \textbf{\bibinfo{volume}{435}},
  \bibinfo{pages}{925} (\bibinfo{year}{2005}).

\bibitem{Paget1977}
\bibinfo{author}{Paget, D.}, \bibinfo{author}{Lampel, G.} \&
  \bibinfo{author}{Sapoval, B.}
\newblock \bibinfo{title}{Low field electron-nuclear spin coupling in gallium
  arsenide under optical pumping conditions}.
\newblock \emph{\bibinfo{journal}{Phys. Rev. B}} \textbf{\bibinfo{volume}{15}},
  \bibinfo{pages}{5780--5796} (\bibinfo{year}{1977}).

\bibitem{Stopa2009}
\bibinfo{author}{Stopa, M.}, \bibinfo{author}{Krich, J.~J.} \&
  \bibinfo{author}{Yacoby, A.}
\newblock \bibinfo{title}{Inhomogeneous nuclear spin flips: Feedback mechanism
  between electronic states in a double quantum dot and the underlying nuclear
  spin bath}.
\newblock \emph{\bibinfo{journal}{Phys. Rev. B}} \textbf{\bibinfo{volume}{81}},
  \bibinfo{pages}{041304} (\bibinfo{year}{2010}).

\bibitem{Maletinsky2007}
\bibinfo{author}{Maletinsky, P.}, \bibinfo{author}{Badolato, A.} \&
  \bibinfo{author}{Imamoglu, A.}
\newblock \bibinfo{title}{Dynamics of quantum dot nuclear spin polarization}.
\newblock \emph{\bibinfo{journal}{Phys. Rev. Lett.}}
  \textbf{\bibinfo{volume}{99}}, \bibinfo{pages}{056804}
  (\bibinfo{year}{2007}).

\bibitem{Reilly2008B}
\bibinfo{author}{Reilly, D.~J.} \emph{et~al.}
\newblock \bibinfo{title}{Exchange control of nuclear spin diffusion in a
  double quantum dot}.
\newblock \emph{\bibinfo{journal}{Phys. Rev. Lett.}}
  \textbf{\bibinfo{volume}{104}}, \bibinfo{pages}{236802}
  (\bibinfo{year}{2010}).

\bibitem{Greilich2009}
\bibinfo{author}{Greilich, A.} \emph{et~al.}
\newblock \bibinfo{title}{Ultrafast optical rotations of electron spins in
  quantum dots}.
\newblock \emph{\bibinfo{journal}{Nature Physics}}
  \textbf{\bibinfo{volume}{5}}, \bibinfo{pages}{262} (\bibinfo{year}{2009}).

\bibitem{Koppens2006}
\bibinfo{author}{{Koppens}, F. H.~L.} \emph{et~al.}
\newblock \bibinfo{title}{Driven coherent oscillations of a single electron
  spin in a quantum dot}.
\newblock \emph{\bibinfo{journal}{Nature}} \textbf{\bibinfo{volume}{442}},
  \bibinfo{pages}{766} (\bibinfo{year}{2006}).

\bibitem{Nowack2007}
\bibinfo{author}{Nowack, K.~C.}, \bibinfo{author}{Koppens, F. H.~L.},
  \bibinfo{author}{Nazarov, Y.~V.} \& \bibinfo{author}{Vandersypen, L. M.~K.}
\newblock \bibinfo{title}{Coherent control of a single electron spin with
  electric fields}.
\newblock \emph{\bibinfo{journal}{Science}} \textbf{\bibinfo{volume}{318}},
  \bibinfo{pages}{1430} (\bibinfo{year}{2007}).

\bibitem{Pioro2008}
\bibinfo{author}{Pioro-Ladriere, M.} \emph{et~al.}
\newblock \bibinfo{title}{Electrically driven single electron spin resonance in
  a slanting zeeman field}.
\newblock \emph{\bibinfo{journal}{Nature Physics}}
  \textbf{\bibinfo{volume}{4}}, \bibinfo{pages}{776--779}
  (\bibinfo{year}{2008}).

\bibitem{Yao2006}
\bibinfo{author}{Yao, W.}, \bibinfo{author}{Liu, R.~B.} \&
  \bibinfo{author}{Sham, L.~J.}
\newblock \bibinfo{title}{Theory of electron spin decoherence by interacting
  nuclear spins in a quantum dot}.
\newblock \emph{\bibinfo{journal}{Phys.\ Rev.\ B.}}
  \textbf{\bibinfo{volume}{74}}, \bibinfo{pages}{195301}
  (\bibinfo{year}{2006}).

\bibitem{Cywinski2009B}
\bibinfo{author}{Cywinski, L.}, \bibinfo{author}{Witzel, W.~M.} \&
  \bibinfo{author}{{Das Sarma}, S.}
\newblock \bibinfo{title}{Electron spin dephasing due to hyperfine interactions
  with a nuclear spin bath}.
\newblock \emph{\bibinfo{journal}{Phys. Rev. Lett.}}
  \textbf{\bibinfo{volume}{102}}, \bibinfo{pages}{057601}
  (\bibinfo{year}{2009}).

\bibitem{Coish2008}
\bibinfo{author}{Coish, W.~A.}, \bibinfo{author}{Fischer, J.} \&
  \bibinfo{author}{Loss, D.}
\newblock \bibinfo{title}{Exponential decay in a spin bath}.
\newblock \emph{\bibinfo{journal}{Phys.\ Rev.\ B.}}
  \textbf{\bibinfo{volume}{77}}, \bibinfo{pages}{125329}
  (\bibinfo{year}{2008}).

\end{thebibliography}

\begin{addendum}
 \item [Acknowledgements] We thank C. Barthel, M. Gullans, B. I. Halperin, J. J. Krich, M. D. Lukin, C. M. Marcus, D. J. Reilly and M. Stopa for discussions. 
We acknowledge funding from ARO/IARPA, the Department of Defense and the National Science Foundation under award number 0653336. This work was performed in part at the Center for Nanoscale Systems (CNS), a member of the National Nanotechnology Infrastructure Network (NNIN), which is supported by the National Science Foundation under NSF award no. ECS-0335765.
 \item[Author Contributions] S.F. fabricated the samples, H.B developed the measurement software, e-beam lithography and sample growth were carried out by D.M. and V.U., respectively. S.F., H.B. and A.Y. planned and performed the experiment, analysed the data and co-wrote the paper.
 
 \item[Correspondence] Correspondence and requests for materials should be addressed to A.Y..
\end{addendum}

\end{document}